\documentclass[a4paper]{article}

\usepackage{INTERSPEECH2019}

\usepackage{amsmath,graphicx}
\usepackage{amssymb}
\usepackage{comment}
\usepackage{color}
\usepackage{cite}

\usepackage[utf8]{inputenc}
\usepackage[linesnumbered,ruled,vlined,commentsnumbered]{algorithm2e}

\SetCommentSty{mycommfont}

\usepackage[hyphens]{url}
\renewcommand{\v}[1]{\ensuremath{\mathbf{#1}}}
\DeclareMathOperator*{\argmax}{arg\,max}

\newcommand{\pdffigure}[3][width=0.7\linewidth]{
	\begin{figure}[tb]
	\begin{center}
    \IfFileExists{./#2.pdf}{
	    \includegraphics[#1]{#2.pdf}
	}{
	    \includegraphics[draft]{#2.pdf}
    }
    \end{center}
    \caption{#3}
	\label{fig:#2}
	\end{figure}
}

\title{End-to-End Neural Speaker Diarization with Permutation-Free Objectives}
\name{
    Yusuke Fujita$^{1,2}$
    \thanks{The first author performed the work while at Center for Language and Speech Processing, Johns Hopkins University as a Visiting Scholar.
    },
     Naoyuki Kanda$^{1}$,
     Shota Horiguchi$^{1}$,
     Kenji Nagamatsu$^{1}$,
     Shinji Watanabe$^{2}$
     }
\address{
    $^{1}$ \text{Hitachi, Ltd. Research \& Development Group, Japan} \\
    $^{2}$ \text{Center for Language and Speech Processing,
            Johns Hopkins University, USA}
}
\email{\{yusuke.fujita.su, naoyuki.kanda.kn, shota.horiguchi.wk, kenji.nagamatsu.dm\}@hitachi.com, shinjiw@ieee.org}

\begin{document}

\maketitle
%
% The total length of the abstract is limited to 200 words
\begin{abstract}
%\begin{comment}
% Kanda's modification
In this paper, we propose a novel end-to-end neural-network-based speaker diarization method.
Unlike most existing methods, our proposed method does not have separate modules for extraction and clustering of speaker representations.
Instead, our model has a single neural network that directly outputs speaker diarization results.
To realize such a model, we formulate the speaker diarization problem as a multi-label classification problem, and introduces a permutation-free objective function to directly minimize diarization errors without being suffered from the speaker-label permutation problem.
Besides its end-to-end simplicity, the proposed method also benefits from being able to explicitly handle overlapping speech during training and inference.
Because of the benefit, our model can be easily trained/adapted with real-recorded multi-speaker conversations just by feeding the corresponding multi-speaker segment labels.
We evaluated the proposed method on simulated speech mixtures.
The proposed method achieved diarization error rate of 12.28\%, while a conventional clustering-based system produced diarization error rate of 28.77\%.
Furthermore, the domain adaptation with real-recorded speech provided 25.6\% relative improvement on the CALLHOME dataset.
Our source code is available online at \url{https://github.com/hitachi-speech/EEND}.

%\end{comment}
\begin{comment}
% 0404 version
In this paper, we propose a novel end-to-end neural speaker diarization method that directly optimizes a diarization-error-oriented objective.
Unlike most existing methods, our proposed method does not have separate modules for extraction and clustering of speaker representations.
These modules are integrated into one neural network that can be jointly optimized to minimize diarization errors.
To realize such an optimization, we formulate the speaker diarization problem as a multi-label classification problem, and introduces permutation-free schemes to solve the speaker-label permutation problem.
Besides its end-to-end simplicity, the proposed method also benefits from being able to explicitly handle overlapping speech during training and inference.
The model can be trained/adapted with real-recorded multi-speaker conversations and the corresponding multi-speaker segment ground truth labeling.
We evaluated the proposed method on simulated speech mixtures.
The proposed method achieved diarization error rate of 12.28\%, while a conventional clustering-based system produced diarization error rate of 28.77\%.
\end{comment}
\end{abstract}

\section{Introduction}
\label{sec:intro}
Speaker diarization is the process of partitioning speech segments according to the speaker identity.
It is an important process for a wide variety of applications such as information retrieval from broadcast news, meetings, and telephone conversations \cite{Tranter2006, Anguera2012}.
It also helps automatic speech recognition performance in multi-speaker conversation scenarios in meetings (ICSI \cite{Janin03, etin2006OverlapIM}, AMI \cite{Renals2008, Kanda2019ICASSP}) and home environments (CHiME-5 \cite{Barker2018, Du2018, Boeddecker2018, Kanda2018, Kanda2019ICASSP}).

Typical speaker diarization systems are based on extraction and clustering of speaker representations \cite{Meignier2010LIUMSA,Shum2013,Sell2014, Senoussaoui2014,Dimitriadis2017, Romero2017, Maciejewski2018CharacterizingPO, Wang2018LSTM}.
The system first extracts speaker representations such as i-vectors \cite{Dehak2011, Shum2013, Sell2014, Maciejewski2018CharacterizingPO}, d-vectors\cite{Wan2018, Wang2018LSTM}, or x-vectors \cite{Snyder2018, Romero2017}.
Then, the speaker representations of short segments are partitioned into speaker clusters.
Various clustering algorithms have been adopted, such as
Gaussian mixture models \cite{Meignier2010LIUMSA, Shum2013}, agglomerative hierarchical clustering \cite{Meignier2010LIUMSA, Sell2014, Romero2017, Maciejewski2018CharacterizingPO},
mean shift \cite{Senoussaoui2014}, k-means \cite{Dimitriadis2017, Wang2018LSTM}, Links \cite{Mansfield2018,Wang2018LSTM}, and spectral clustering \cite{Wang2018LSTM}.
These clustering-based diarization methods have shown to be effective in various datasets (see the DIHARD challenge 2018 activities, e.g.,  \cite{Sell2018dihard,Diez2018,Sun2018}).

However, there are two problems in the clustering-based method.
Firstly, the clustering-based method implicitly assumes one speaker per segment, so it is difficult to deal with speaker-overlapping speech.
Secondly, it cannot be optimized to minimize diarization errors directly because the clustering is performed in an unsupervised manner.

To deal with speaker-overlapping speech, a neural network based source separation model was recently proposed \cite{Neumann2019}. The model separates one speaker's time-frequency mask in one iteration, and separates another speaker's mask in another iteration. 
Utilizing the source separation technique, speaker diarization is realized even in overlapping speech.
However, their source separation training objective does not necessarily minimize diarization errors.
%However, their model is not optimized in terms of diarization errors but rather signal reconstruction errors.
Aiming at the speaker diarization problem, it will be better to use a diarization error-oriented objective function.
Moreover, there is another drawback in their method that real multi-speaker recordings cannot be used for training, because their model requires clean, non-overlapping reference speech for training.

For the optimization based on diarization errors, a fully supervised diarization method has been proposed \cite{Zhang2018}.
This method formulates the speaker diarization problem based on a factored probabilistic model, which consists of modules for speaker change, speaker assignment and feature generation.
However, in their method, the speaker-change model assumes one speaker for each segment, which hinders the application of the method for speaker-overlapping speech.

In this paper, 
we propose a novel end-to-end neural network-based speaker diarization model (EEND). In contrast to previous techniques, the EEND can both deal with overlapping speech as well as be trained directly to minimize diarization errors.
Given an audio recording with utterances by multiple speakers, our recurrent neural network estimates joint speech activities of all speakers frame-by-frame.
This model is categorized as multi-label classification, similar to the well-known method in sound event detection (SED) \cite{Adavanne2017}.
Unlike SED, the output speaker labels are ambiguous in the training stage, i.e. not corresponding to any fixed class, which is known in the source separation research field as a permutation problem.
To solve the problem, we introduce a permutation-free scheme \cite{Hershey2016, Yu2017} into the training objective function.
The model is trained in an end-to-end fashion using the objective function that provides minimal diarization errors.

The EEND has various advantages over the conventional methods.
Firstly, the EEND can explicitly handle overlapping speech by simply feeding overlapping speech as input during training and inference.
Secondly, the EEND does not require separate modules for speech activity detection, speaker identification, source separation, or clustering.
The proposed model integrates their functionality into a single neural network.
Thirdly, unlike the source separation model, the EEND does not require clean, non-overlapping speech for training the model with synthetic conversational mixtures.
This enables the use of domain adaptation using real overlapping speech conversations.
\section{Proposed Method}

\subsection{Neural probabilistic model of speaker diarization}

Speaker diarization is the process of partitioning speech segments
according to the speaker identity.
In other words, speaker diarization determines ``who spoke when.''
We formulate the speaker diarization task as a multi-label classification problem.
It can be formulated as follows:

Given an observation sequence $X = (\v{x}_t \in \mathbb{R}^{F} \mid t=1,\cdots,T)$ from an audio signal, estimate the speaker label sequence $Y = (\v{y}_t \mid t=1,\cdots,T)$.
Here, $\v{x}_t$ is a $F$-dimensional observation feature vector at time index $t$.
Speaker label $\v{y}_t = [y_{t,c} \in \{0,1\} \mid c=1, \cdots, C]$ denotes a joint activity for multiple ($C$) speakers at time index $t$.
For example, $y_{t,c} = 1$ and $y_{t,c'} = 1$ represent an overlap situation of both speakers $c$ and $c'$ being present at time index $t$.
Thus, determining $Y$ is a sufficient condition to determine the speaker diarization information.

The most probable speaker label sequence $\hat{Y}$ is estimated
among all possible speaker label sequences $\mathcal{Y}$, as follows:
\begin{equation}
\hat{Y} = \argmax_{Y \in \mathcal{Y}} P(Y|X).
\end{equation}
$P(Y|X)$ can be factorized using conditional independence assumption as follows:
\begin{align}
    P(Y|X) &= \prod_{t} P(\v{y}_t | \v{y}_1, \cdots \v{y}_{t-1}, X), \\
    &\approx \prod_t P(\v{y}_t| X) \approx \prod_t \prod_c P(y_{t,c}| X).
\end{align}
Here, we assume the frame-wise posterior is conditioned
on all inputs, and each speaker is present independently.

The frame-wise posterior $P(y_{t,c}| X)$ is modeled with bi-directional long short-term memory (BLSTM), as follows:
\begin{align}
    \v{h}^{(1)}_t &= \mathrm{BLSTM}_t(\v{x}_1, \cdots, \v{x}_T) \in \mathbb{R}^{2H},\\
    \v{h}^{(p)}_t &= \mathrm{BLSTM}_t(\v{h}^{(p-1)}_1, \cdots, \v{h}^{(p-1)}_T) \  (2 \le  p \le P), \label{eq:hidden} \\
    \v{z}_t  &= \sigma(\mathrm{Linear}(\v{h}^{(P)}_t)) \in (0,1)^C,    \label{eq:out}
\end{align}
where $\mathrm{BLSTM}_t(\cdot)$ is a BLSTM layer which accepts an input sequence and outputs $2H$-dimensional hidden activations $\v{h}_t^{(p)}$ at time index $t$.\footnote{It is a concatenated vector of $H$-dimensional forward and backward LSTM outputs.}
We use $P$-layer stacked BLSTMs.
%$\mathrm{Linear}(\cdot)$ is a linear layer which converts a  $2H$-dimensional hidden activations to $C$-dimensional vector, and $\mathrm{Sigmoid}(\cdot)$ is the element-wise sigmoid function. $\v{z}_t$ is a $C$-dimensional output vector which represents the frame-wise posteriors $P(y_{t,c}| X)$.
The frame-wise posteriors $\mathbf{z}_t$ is calculated from $\mathbf{h}_t^{(P)}$ using a fully-connected layer $\mathrm{Linear}:\mathbb{R}^{2H}\rightarrow\mathbb{R}^{C}$ and the element-wise sigmoid function $\sigma\left(\cdot\right)$.

The difficulty on training of the model described above is that the model have to deal with the speaker permutations: changing an order of speakers within a correct label sequence is also regarded as correct. An example of the permutations in a two-speaker case is shown in Fig. \ref{fig:trainingflow}.
In this paper, we call this the label ambiguity. This label ambiguity obstructs the training of the neural network when we just use a standard binary cross entropy loss function.

\pdffigure[width=0.92\linewidth]{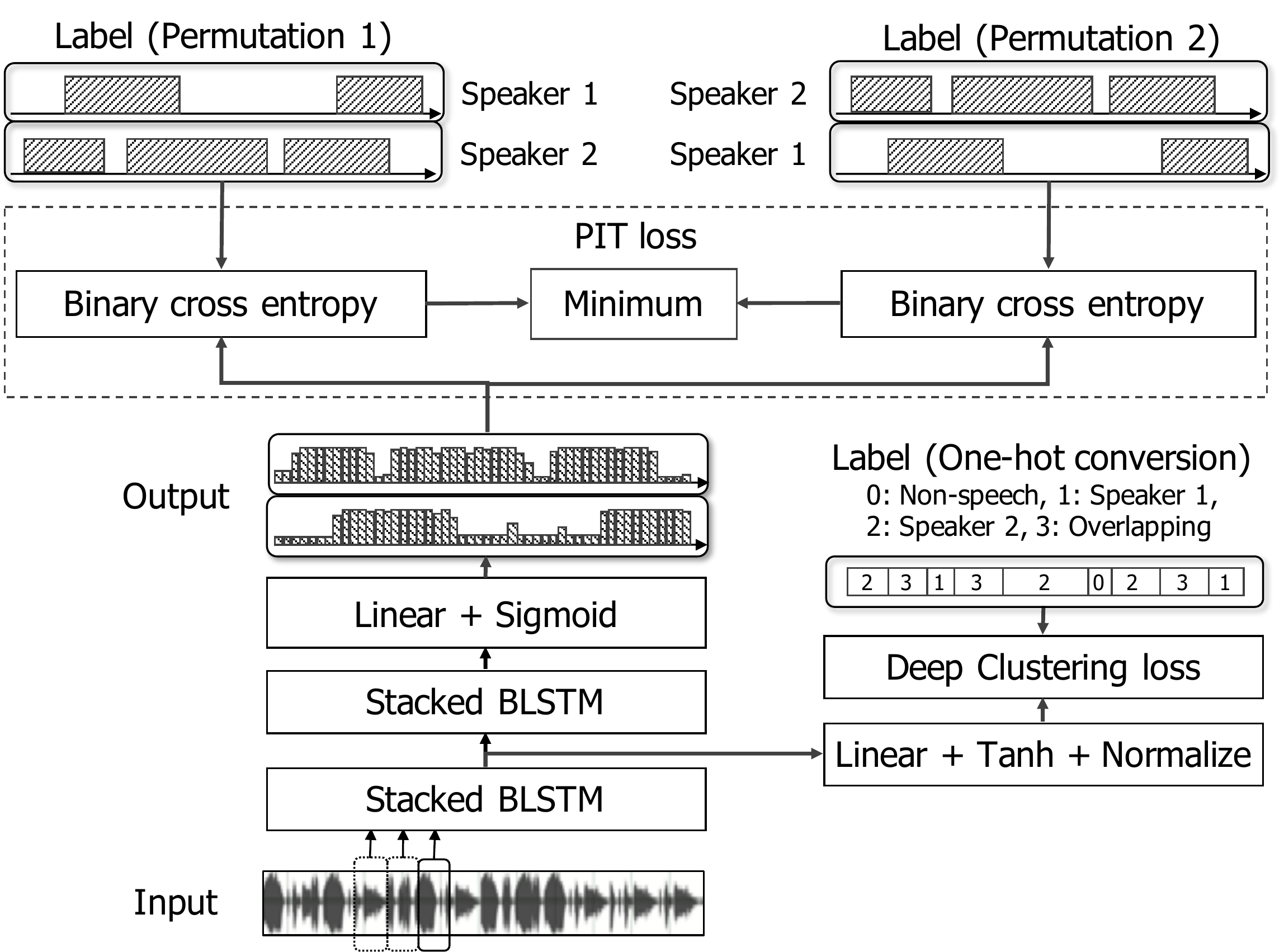}{Two-speaker end-to-end neural speaker diarization (EEND) model trained with the PIT loss and the DPCL loss}

To cope with the label ambiguity problem, we introduce two permutation-free loss functions as shown in Fig. \ref{fig:trainingflow}.
The first loss function is the permutation-invariant training (PIT) loss function, which is used for considering all the permutations of ground-truth speaker labels. The second loss function is the Deep Clustering (DPCL) loss function, which is used for encouraging hidden activations of the network to be speaker-discriminative representations.
Note that the use of multi-label classification model is similar to the well-known SED method \cite{Adavanne2017}.
However, in contrast to the SED, the multi-label classification model used for speaker diarization suffers from the label ambiguity problem.
Our contribution is introducing two permutation-free loss functions to cope with the label ambiguity problem.

\subsection{Permutation-invariant training loss}

The neural network is trained to minimize the error between the output $\v{z}_t$ predicted in Eq. \ref{eq:out} and the ground-truth speaker label $\v{l}_t$.
Considering that the speaker label has ambiguity of their permutations, we introduce the permutation-free scheme \cite{Hershey2016, Yu2017}. More specifically, we utilize the utterance-level permutation-invariant training (PIT) criterion \cite{Kolbak2017} in the proposed method.
We apply the PIT criterion on time sequence of speaker labels instead of time-frequency mask used in \cite{Kolbak2017}.
The PIT loss function is written as follows:
\begin{equation}\label{eq:pf}
    J^{\text{PIT}} = \frac{1}{TC} \min_{\phi \in \mathrm{perm}(C)} \sum_t \mathrm{BCE}(\v{l}_t^\phi, \v{z}_t),
\end{equation}
where $\mathrm{perm}(C)$ is a set of all the possible permutation of ($1,\dots,C$),
and $\v{l}_t^\phi$ is the $\phi$-th permutation of the ground-truth speaker label, $\mathrm{BCE}(\cdot, \cdot)$ is the binary cross entropy function between the label and the output.
%\begin{equation}
%\mathrm{BCE}(\v{l}_t,\v{z}_t) = \sum_c - l_{t,c}\log(z_{t,c}) - (1-l_{t,c})\log(1-z_{t,c}).
%\end{equation}

\subsection{Deep Clustering loss}

Assuming that the neural network extracts speaker representation in lower layers and then performs segmentation using higher layers, the middle layer activations can be regarded as the speaker representation.
Therefore, we introduce a speaker representation learning criterion on the middle layer activations.

Here, the $q$-th layer activations $\v{h}^{(q)}_t$ obtained from Eq.~\ref{eq:hidden} are transformed into normalized  $D$-dimensional embedding $\v{v}_t$ as follows:
\begin{align}
    \v{v}_t &= \mathrm{Normalize}(\mathrm{Tanh}(\mathrm{Linear}(\v{h}^{(q)}_t))) \in \mathbb{R}^D \label{eq:dc3},
\end{align}
where $\mathrm{Tanh}(\cdot)$ is the element-wise hyperbolic tangent function and $\mathrm{Normalize}(\cdot)$ is the L2 normalization function.
We apply the Deep Clustering (DPCL) loss function \cite{Hershey2016} so that the embeddings are partitioned into speaker-dependent clusters as well as overlapping and non-speech clusters.
For example in a two-speaker case, we generate four clusters (Non-speech, Speaker 1, Speaker 2, and Overlapping) as shown in Fig. \ref{fig:trainingflow}.

DPCL loss function \cite{Hershey2016} is used as follows:
\begin{align}
    J^\text{DC} &= \|VV^\top - L'L'^\top\|_F^2,
\end{align}
where $V = [\v{v}_1,\cdots,\v{v}_T]^\top$, and $L' \in \mathbb{R}^{T \times 2^C}$ is a matrix for each row represents one-hot vector converted from $\v{l}_t$ where those elements are in the power set of speakers.
$\|\cdot\|_F$ is the Frobenius norm.
The loss function encourages the two embeddings at different time indices to be close together if they are in the same cluster and far away if they are in different clusters.

Then we use multi-objective training introducing a mixing parameter $\alpha$:
\begin{align}
    J^{\text{MULTI}} = (1 - \alpha) J^{\text{PIT}} + \alpha J^{\text{DC}}.
\end{align}
Thus, we derive end-to-end neural speaker diarization with the above permutation-free objective.

\section{Experiments}

\subsection{Data}
\label{sec:data}

%To validate the performance of the proposed neural network, we generate simulated speech mixtures.
This paper mainly conducted our experiments with simulated speech mixtures to verify the effectiveness of the proposed method for controlled overlap situations.
Each mixture is simulated by Algorithm \ref{alg:mixture_simulation}.
Unlike the existing mixture simulation for source separation studies \cite{Hershey2016}, we consider a diarization-style mixture simulation: each speech mixture should have dozens of utterances per speaker with reasonable silence intervals between utterances.
The silence intervals are controlled by the average interval $\beta$. Larger $\beta$ values generate speech with less overlap. 
We show performance for differing overlap ratio controlled by $\beta$ in the result section Sec.\ref{sec:result}.

\begin{algorithm}[t]
	\SetAlgoLined
	\DontPrintSemicolon
	\caption{Mixture simulation.}
	\label{alg:mixture_simulation}
	\SetAlgoVlined
	\SetKwInOut{Input}{Input}
	\SetKw{In}{in}
	\Input{
	    {{$\mathcal{S,N,I,R}$} \tcp*{Set of speakers, noises, RIRs and SNRs}}\\
	    {{$\mathcal{U} = \{U_s\}_{s \in \mathcal{S}}$} \tcp*{Set of utterance lists}}
	    {{$N_\text{spk}$} \tcp*{\#speakers per mixture}}
	    {{$N_\text{umax},N_\text{umin}$} \tcp*{Max. and min. \#utterances per speaker}}
	    {{$\beta$} \tcp*{average interval}}
	}
	\SetKwInOut{Output}{Output}
	\Output{$\mathbf{y}$\tcp*{mixture}}
	\BlankLine
	Sample a set of $N_\text{spk}$ speakers $\mathcal{S'}$ from $\mathcal{S}$\\
		$\mathcal{X}\leftarrow\emptyset$\tcp*{Set of $N_\text{spk}$ speakers' signals}
		\ForAll{$s\in\mathcal{S'}$}{
			$\mathbf{x}_s\leftarrow\emptyset$\tcp*{Concatenated signal}
			Sample $\mathbf{i}$ from $\mathcal{I}$\tcp*{RIR}
			Sample $N_u$ from $\left\{N_\text{umin},\dots,N_\text{umax}\right\}$%\tcp*{\#utterances per speaker}
			
			\For{$u=1$ \rm{to} $N_u$}{
				Sample $d\sim\frac{1}{\beta}\exp\left(-\frac{d}{\beta}\right)$\tcp*{Interval}
				$\mathbf{x}_s\leftarrow\mathbf{x}_s\oplus\mathbf{0}^{\left(d\right)}\oplus U_s\left[u\right]\ast\mathbf{i}$
			}
			$\mathcal{X}.\mathsf{add}\left(\mathbf{x}_s\right)$\\
		}
		$L_\mathrm{max}=\max_{\mathbf{x}\in\mathcal{X}}\lvert\mathbf{x}\rvert$\\
		$\mathbf{y}\leftarrow\sum_{\mathbf{x}\in\mathcal{X}}\left(\mathbf{x}\oplus\mathbf{0}^{\left(L_\mathrm{max}-\lvert\mathbf{x}\rvert\right)}\right)$\\
		Sample $\mathbf{n}$ from $\mathcal{N}$\tcp*{Background noise}
		Sample $r$ from $\mathcal{R}$\tcp*{SNR}
		Determine a mixing scale $p$ from $r,\mathbf{y},$ and $\mathbf{n}$\\
		$\mathbf{n}'\leftarrow$ repeat $\mathbf{n}$ until reach the length of $\mathbf{y}$\\
		$\mathbf{y}\leftarrow\mathbf{y}+p\cdot\mathbf{n}'$\\
\end{algorithm}

The set of utterances used for the simulation is comprised of Switchboard-2 (Phase I, II, III), Switchboard Cellular (Part 1, Part2), and NIST Speaker Recognition Evaluation datasets (2004, 2005, 2006, 2008). All recordings are telephone speech sampled at 8 kHz.
Total number of speakers in these corpora is 6,381.
We split them into 5,743 speakers for the training set and 638 speakers for the test set.
Since there is no time annotations in these corpora, we extract utterances using speech activity detection (SAD) based on time-delay neural networks and statistics pooling\footnote{The SAD model: \url{http://kaldi-asr.org/models/m4}}. This data preparation and SAD is performed using Kaldi speech recognition toolkit \cite{Povey_ASRU2011}.

The set of background noises is from MUSAN corpus \cite{Snyder2015}. We used 37 recordings which are annotated as ``background'' noises.
%Since the noises are sampled at 16 kHz, we downsampled them to 8 kHz.
The set of room impulse responses (RIRs) is the Simulated Room Impulse Response Database used in \cite{Ko2017}. The total number of RIRs is 10,000.
The SNR values are sampled from 10, 15, and 20 dBs.
We generated two-speaker mixtures for each speaker have 20-40 utterances ($N_{\text{spk}} = 2, N_{\text{umin}}=20, N_{\text{umax}}=40$).
We used differing number of mixtures for the training set, and 500 mixtures for the test set.

\subsection{Experimental setup}
We extracted 23-dimensional log-Mel-filterbank features with 25 ms frame length and 10 ms frame shift.
Each features are concatenated with those from the previous 7 frames and subsequent 7 frames.
To deal with a long audio sequence in our BLSTM, we subsampled the concatenated features by a factor of 10.

For our neural network, we used 5-layer BLSTM with 256 hidden units in each layer. For the DPCL loss, we used the second layer of BLSTM outputs to form 256-dimensional embedding.
We used the Adam \cite{Kingma2014} optimizer with initial learning rate of $10^{-3}$.
The batch size was 10.
The number of training epoch was 20.
Our implementation was based on Chainer \cite{chainer_learningsys2015}.

Because the output of the neural network is a probability of speech activity for each speaker, a threshold is required to obtain the decision of speech activity for each frame. We set the threshold to 0.5 for the evaluation on simulated speech mixtures.
Furthermore, we apply 11-frame median filtering to prevent the production of unreasonably short segments.

\subsection{Performance metric}

We evaluated the proposed method with diarization error rate (DER) \cite{NISTRT09}.
In many prior studies, DER had not included miss or false alarm errors due to using oracle speech/non-speech labels.
Overlapping speech segments had also been excluded from the evaluation.
For our DER computation we evaluated all errors, including both non-speech and overlapping speech segments, because the proposed method includes both speech activity detection and overlapping speech detection functionality.
As is typical, we use a collar tolerance of 250 ms around both the start and end of each segment.

\subsection{Baseline system}
We compared the proposed method with the two conventional clustering-based systems \cite{Sell2018dihard}.
The i-vector system and the x-vector system were created using the Kaldi CALLHOME diarization recipe\footnote{\url{https://github.com/kaldi-asr/kaldi/tree/master/egs/callhome_diarization}}.
To evaluate  non-speech segments, we used speech segments extracted by SAD as described in Sec. \ref{sec:data}.

\subsection{Results}
\label{sec:result}

\begin{table}[t]
\caption{Effect of loss functions evaluated on simulated speech generated with $\beta=2$. The models are trained using 10,000 mixtures genarated with $\beta=2$.}
\label{tab:loss}
\centering
\begin{tabular}{ccc} \hline
PIT loss & DPCL loss & DER (\%) \\ \hline
- & - & 41.74 \\
\checkmark & -  & 25.14 \\
\checkmark & \checkmark  & 23.79 \\ \hline
\end{tabular}
\end{table}

\begin{table}[t]
\caption{Effect of the number of training mixtures evaluated on simulated speech generated with $\beta=2$.
The models are trained with $\beta=2$.}
\label{tab:sample}
\centering
\begin{tabular}{cc} \hline
Number of training mixtures &	DER(\%) \\ \hline
10,000	& 23.79 \\
20,000 &	14.66 \\
100,000 &	12.28 \\ \hline
\end{tabular}
\end{table}

%\pdffigure{chart1}{Effect of the number of training mixtures evaluated on simulated speech generated with $\beta=2$.
%The models are trained with $\beta=2$.}

\begin{table}[t]
\caption{Detailed DERs (\%) evaluated on simulated speech generated with the $\beta = 2$.  MI, FA and CF denote miss, false alarm and confusion error rates, respectively. The proposed model is trained using 100,000 mixtures generated with $\beta=2$.}
\label{tab:detail}
\centering
\begin{tabular}{ccccc} \hline
Method & DER	& MI & FA & CF \\ \hline
i-vector & 33.74 & 25.82 & {\bf1.05} & 6.88 \\
x-vector &	28.77& 25.82 & {\bf1.05} &	{\bf1.90} \\
EEND (proposed) &	{\bf12.28} &	\bf{4.47}	& 5.20	&  2.61  \\ \hline
\end{tabular}
\end{table}

\begin{table}[t]
\caption{DERs (\%) on different overlapping conditions.
For the evaluation on simulated mixtures, the proposed model is trained using 100,000 mixtures generated with $\beta=2$. For the evaluation on the CALLHOME dataset, the proposed model is trained with 26,712 telephone recordings. The DER without the domain adaptation is shown in the parenthesis.}
\label{tab:overlap}
\centering
\begin{tabular}{c|ccc|c} \hline
Evaluation set & \multicolumn{3}{c|}{Simulated mixtures} & CALLHOME \\
$\beta$ & 2 & 3 & 5 & - \\
overlap ratio (\%) & 27.3 & 19.1 & 11.1 & 11.8 \\ \hline \hline
i-vector & 33.74 & 30.43 & 25.96 & 12.10\\
x-vector &	28.77& 24.46 &	19.78 & \bf{11.53} \\
EEND &	\bf{12.28} & \bf{14.36} & {\bf19.69}  &23.07 (31.01) \\ \hline
\end{tabular}
%\vspace{1em}
%\caption{DER(\%) with adaptation to the CALLHOME dataset.
%The proposed model without adaptation is trained using 100,000 mixtures generated with $\beta=2$.
%}
%\label{tab:adaptation}
%\centering
%\begin{tabular}{ccc} \hline
%Method & DER \\ \hline
%proposed & 34.33 \\
%proposed+adaptation & 26.90 \\ \hline
%\end{tabular}
\end{table}

We evaluated the effect of the proposed loss functions.
Without PIT loss, we used binary cross entropy loss with the fixed permutation\footnote{We sorted the speaker names in a lexical order to obtain the fixed permutation.}. With PIT and DPCL losses, we set the mixing parameter $\alpha = 0.5$.
The results are shown in Table \ref{tab:loss}.
It is observed that PIT loss is essential for training of our neural network. It also demonstrates that DPCL loss helps improve performance.

%\subsection{Effect of training data size}

The comparison with different numbers of training mixtures is shown in Table \ref{tab:sample}.
It is observed that increasing the number of training samples improves the performance.
Because our proposed method can be trained with any speech mixture with corresponding time annotations, it is possible to utilize large scale speech corpora for improving robustness of the system.

We compared the proposed method with the baseline systems using the simulated speech mixtures.
The results are shown in Table \ref{tab:detail}.
It is observed that miss rate is dominant in the DER of the baseline systems, due to the lack of capability for overlapping speech.
In contrast, the proposed method achieved significantly low miss rate.
The results indicate that the proposed method successfully detects overlapping segments as well as single-speaker segments and silence segments. 
Regarding the confusion error rate, the proposed method is better than the i-vector system, while it is worse than the x-vector system.
For reducing the confusion errors, it is possible to use data augmentation for learning noise and speaker variations, which is utilized in the x-vector system.

To investigate the robustness to variable conditions, we evaluated different overlap ratio controlled by the average interval $\beta$. Larger $\beta$ values generate less overlapping speech.
The results with different overlapping ratios are shown in Table \ref{tab:overlap}.
The baseline systems show better performance on less overlapping speech as expected.
However, the proposed method unexpectedly showed degraded performance on less overlapping speech.
The result suggests that the network had overfit to the specific overlap ratio: 27.3\%.
Investigation with various overlap ratio settings of training data is among our future work.

In addition,
we evaluated the proposed method on real telephone conversations using the CALLHOME dataset.
We split two-speaker recordings from the CALLHOME dataset into two subsets: an adaptation set of 155 recordings and a test set of 148 recordings.
Our neural network was trained with a set of 26,172 two-speaker recordings from telephone speech recordings as described in Sec.\ref{sec:data}. The overlap ratio of the training data was 5.8\%.
Then, it was retrained with the adaptation set.
For this retraining, we used the Adam optimizer with initial learning rate of $10^{-6}$ and ran 5 epochs.
For the postprocessing, we adjusted the threshold to 0.6 so that the DER of the adaptation set has the minimum value.
Table \ref{tab:overlap} shows the DERs evaluated on the CALLHOME test set.
Unfortunately, the proposed method produces worse DER than the baseline systems.
This is likely because our training set has very different overlap ratio (5.8\%) from the CALLHOME test set (11.8\%).
To reduce this condition mismatch, we tried domain adaptation.
The result showed a significant DER reduction.
The relative improvement introduced by domain adaptation was 25.6\%.
Although the DER of the proposed method was still behind those of the baseline systems, we expect it will be much improved by developing better simulation techniques of training data or just by feeding more real data, as was suggested by the result with simulated data.
We will address these directions in our future work.

%\subsection{Effect of domain adaptation}

%\subsection{Error analysis on overlapping segment}

%\pdffigure{visualize}{Visualization of overlapping speech}

\section{Conclusion}

We proposed an end-to-end neural speaker diarization method that is directly optimized with a diarization-error-oriented objective.
The experimental results show that the proposed method outperforms conventional clustering-based methods evaluated on simulated speech mixtures.
Furthermore, domain adaptation with real speech data achieved a significant DER reduction on the CALLHOME dataset.

\section{Acknowledgements}

We would like to thank Matthew Maciejewski and Xuankai Chang for their
comments that greatly improved the manuscript.

\bibliographystyle{IEEEtran}

\bibliography{refs}

\end{document}